\documentstyle{mn}
\def\Msun{\ifmmode {\rm M}_{\odot} \else M$_{\odot}$\fi}
\def\Mdot{\ifmmode {\rm \dot M} \else ${\rm \dot M}$\fi}
\newcommand{\ltorder}{\mathrel{<\kern-1.0em\lower0.9ex\hbox{$\sim$}}}
\newcommand{\gtorder}{\mathrel{>\kern-1.0em\lower0.9ex\hbox{$\sim$}}}
\input psfig
\title{Thermal Instability and Evaporation of Accretion Disc Atmospheres}
\author[de Kool \& Wickramasinghe]{Martijn de Kool and Dayal Wickramasinghe \\
Astrophysical Theory Centre \thanks{Operated jointly by the
Research School of Astronomy and Astrophysics and the School of Mathematical
Sciences, The Australian National University, ACT 0200, Australia}, 
Australian National
University, ACT 0200, Australia}        
\begin{document}
\maketitle
\begin{abstract}
We investigate the vertical structure of the outer
layers of accretion discs
in which the local viscous energy dissipation rate scales with the pressure as
for standard Shakura-Sunyaev discs. It has been pointed
out by several authors that a thermal instability occurs in the outer
layers of such discs when the gas pressure drops below a certain
value. When the density becomes too low thermal equilibrium can no
longer be maintained and the gas heats up, forming a hot corona or
possibly a wind. To assess the importance of this effect we estimate
the pressure and temperature at which this instability will occur,
where the instability point lies with respect to the total vertical
disc structure, and whether the instability is likely to be important
for the disc as a whole. The main difference between our work and
earlier estimates lies in a more detailed treatment of the heating and
cooling processes and the inclusion of the effects of an external
radiation field.

By solving for the accretion disc vertical structure using the
grey two-stream approximation instead of the usual diffusion
approximation for the radiative transfer, we first show that the
thermal structure of the optically thin outer layers is in first 
approximation independent from radiative transfer effects, and 
follows the thermal equilibrium curve for optically thin plasma's in
the pressure-temperature plane. We then calculate
the thermal structure using the detailed photoionisation code
MAPPINGS, which includes much more accurate heating and cooling 
physics than the mean opacity used in the vertical structure calculations. 
This approach also allows a
straightforward inclusion of the effects of an external radiation
field from the centre of the accretion flow.

We apply our method to cataclysmic variable (CV), and stellar mass black
hole discs, and show that evaporation due to the thermal instability
can be important under a variety of conditions. In the case of CVs,
radiative heating by photons emanating from the 
boundary layer can increase the evaporation rate  
significantly over the non-irradiated case, but for
steady state CV discs the evaporation by the mechanism considered here
is still not sufficient to evaporate the entire disc. 
It may become important however in non-steady discs in dwarf novae
if the accretion heated white dwarf plays a role in irradiating the disc 
after an outburst.
In the case of Black Hole Soft X-ray Transients, the evaporation can have a
significant effect on the structure of the outer regions of the disc, 
resulting in mass
loss rates comparable to the local mass accretion rate through the disc
for $\Mdot \ltorder 10^{16}$ g s$^{-1}$ for a $10$ solar mass black hole. 
Accretion in such systems could 
therefore proceed mainly from a hot thick disc formed by evaporation
from the outer regions of the thin disc. The evaporation can be quenched by 
Compton cooling only for mass transfer rates of $\Mdot \gtorder
10^{18}$ g s$^{-1}$ and low values of the viscous heating parameter $\alpha$.

\end{abstract}


\section{Introduction}
The vertical structure of accretion discs, as based on the equations
of hydrostatic equilibrium, energy transport and detailed
opacities and equation of state has been studied extensively in the
past (e.g. Mineshige \& Osaki 1983, Smak 1982, Meyer \& Meyer Hofmeister 1982, 
Canizzo \& Wheeler 1984, Shaviv \& Wehrse 1986, Mineshige \& Wood 1990). 
Most of these works assume that the heating prescription for 
vertically averaged disc structures originally given by Shakura \& Sunyaev
(1972), $H \propto \alpha {\rm P}$, is also valid
locally. As discussed earlier in Shaviv \& Wehrse 1986 (hereafter SW), Adam et
al. 1988 (hereafter ASSW), Czerny \& King 1989
 and Shaviv, Wickramasinghe \& Wehrse 1998
(hereafter SWW) this assumption must lead to a thermal 
instability in the outer, low-density layers of the disc.
The underlying mechanism is as follows. The cooling rate at a given 
temperature of a low density optically thin gas 
is roughly proportional to the density squared, whereas 
the viscous heating rate is
proportional to the density. Thus, when the density in the disc
decreases outwards, the cooling decreases faster than the
heating. As long as the
cooling rate at constant pressure increases with temperature, 
the equilibrium temperature rises with decreasing density. More realistic
cooling curves however (e.g. Raymond et al. 1976) have a maximum in
the cooling rate at a temperature
between $10^4$ and $10^5$ K, after which the cooling rate decreases
with increasing temperature. This leads to runaway heating.

In this paper we investigate this instability further, and discuss the
likely consequences for the disc structure. ASSW
have argued that this instability can lead to the formation
of a stable ``corona'' with a temperature of a few times $10^5$ K over
a large fraction of the disc surface, and SWW argue that it could lead 
to mass loss from the surface of the disc that can become comparable 
to the total mass accretion rate. 

These conclusions depend strongly on the opacities used in these
calculations, since the opacity essentially prescribes the cooling rate. 
SWW considered power law opacities, and in their applications to
Black Hole Soft X-ray Transients (BHSXTs),
took only bremsstrahlung emission and absorption
into account, which severely underestimates  the cooling rate in
the optically thin outer disc layers when they are cooler than $10^7 -
10^8$ K. ASSW used relatively old opacity
tables, and as is discussed in more detail below,
used the Rosseland mean absorption coefficient in the thermal
equilibrium equation rather than the Planck mean (see also Hubeny 1990). 
These two means can easily differ by a factor of 
$10^3$, so that this assumption has a major effect on the results.
Additionally, these papers did not consider the effect of the
irradiation of the outer disc layers by a much harder, but diluted
continuum from the central parts of the accretion disc or the
accreting compact object. 

We therefore reinvestigate the structure of the optically thin outer
layers of accretion discs. The outline of this paper is as follows.
First, we present some vertical structure models based on the two
stream approximation. These show that
the thermal structure of the outer layers is essentially determined
by the thermal equilibrium curve in the pressure - temperature plane.
We then argue (see also Mineshige \& Wood 1990)
that the description of heating and cooling processes 
in terms of mean opacities (as used in the set of accretion disc
vertical structure equations) is not accurate enough to properly
describe the thermal equilibrium, and present a set of detailed
calculations of thermal equilibrium curves in the $P - T$ plane obtained with
the photoionisation equilibrium code MAPPINGS. Based on these curves, we then
discuss the consequences of the thermal instability for the accretion
disc. The results are applied to Cataclysmic variable (CV),
 and stellar mass black hole discs. 
 
\section{Method}

\subsection{Vertical Structure Equations \label{eqsec}}

To solve for the detailed vertical disc structure we require a set of
equations for the hydrostatic equilibrium and energy generation and
transport. Since this paper deals mainly with processes in the outer,
optically thin layers of accretion discs we neglect convective energy
transport. 

The hydrostatic equation is

\begin{equation} 
{{d P}\over{dz}} = -\Omega ^2 z \rho (P,T) + {{\chi_R}\over c}
 F \label{hstat}
\end{equation}

where $P$ is the gas pressure, and
$\Omega$ the Keplerian angular velocity of the disc at the radius
considered, $\chi_R$ is the Rosseland mean opacity and $F$ the
radiation flux. Our equation of state 
\begin{equation} 
\rho = \rho (P,T)
\label{eqstate} 
\end{equation}
includes the effects of ionization of hydrogen and helium, and is
computed together with the mean opacities.

Most detailed vertical structure models calculate the radiative energy
transport in the diffusion approximation (Meyer \& Meyer-Hofmeister 1982,
Canizzo \& Wheeler 1984), which is clearly not appropriate for the optically
thin outer regions of the disc that are the main emphasis 
of this work.  The best way to solve the problem would be to treat the
entire disc as a stellar atmosphere (see e.g. Hubeny 1990, Hubeny \&
Hubeny 1997) with a full
treatment of the angle- and frequency dependent radiation field, but in
practice this is not straightforward, and also not necessary if one is
just interested in the global behaviour of the solutions and not in
the detailed line profiles. We therefore 
make two approximations. The frequency dependence is eliminated by
using only frequency integrated quantities and appropriate mean
opacities, and
the angle dependence is simplified by considering only an ingoing and
an outgoing direction. This leads to the so-called grey two-stream
approximation. 
Although approximative, this method still allows for a natural 
transition between optically thick and optically thin regions. 

We base ourselves on the grey two-stream formalism,
in which the direction of the outgoing and ingoing beam are taken 
to travel at an angle $\theta$ with respect to the normal of the
surface , with $\cos \theta = {1\over{\sqrt 3}}$ and $ -{1\over{\sqrt
3}}$ respectively. In a slightly different form (taking 
$\cos \theta = 1, -1$), this
formalism has been used in several papers calculating accretion
disc vertical structure (SW, ASSW).

We first define the following quantities. The frequency integrated
intensities along the outgoing and ingoing  stream will
be called ($I^+$) and ($I^-$). In terms of these two quantities, 
the mean intensity $J$ and flux $F$ are defined as
\begin{equation}J = {1 \over 2} ( I^+ + I^- )\end{equation}
\begin{equation}F = {{2 \pi}\over{\sqrt 3}} (I^+ - I^-) \end{equation}
The radiative transfer equations for ($I^+$) and ($I^-$) are
\begin{equation}{1 \over {\sqrt 3}}{{d I^+}\over{dz}} =
-\chi \rho I^+ + j = -\chi \rho I^+ + \kappa \rho B(T) 
\end{equation}
\begin{equation}-{1 \over {\sqrt 3}}{{d I^-}\over{dz}} =
-\chi \rho I^- + j = -\chi \rho I^- + \kappa \rho B(T)
\end{equation}
where $\chi$ and $\kappa$ are the mean opacity and the absorption
coefficient (per gram), $j$ is the volume emissivity, and $B(T)$ the
frequency integrated Planck function. The second equality
follows from the assumption of LTE. 

By adding and subtracting the transfer equations for $I^+$ and
$I^-$ we derive the folowing equations for $J$ and $F$ (similar to the  
first and second moments of the full radiative transfer equation)
\begin{equation} 
{dJ \over dz} = - { {3 \chi_R \rho} \over {4 \pi}} F \label{dJdz}
\end{equation}  
\begin{equation} 
{dF \over dz} = 4 \pi \kappa_P \rho \left( B(T) - J \right) 
\label{dFdz}
\end{equation}
where $\chi_R$ is the Rosseland mean opacity, and
 $ \kappa_P $ the Planck mean absorption coefficient (both per gram). 
Equation \ref{dFdz} implicitly assumes that the spectral shape of
the mean intensity $J$ is not too different from the Planck function
at the local temperature $T$. This could be a problem if the
temperature in the optically thin outer disc layers would be very
different from the disc effective temperature (corona-like), but this
situation is not encountered in our solutions.  

From the definitions of $J$ and $F$, and the condition that
there is no incoming radiation at the outer boundary \hbox{($I^- = 0$)}, 
and the symmetry condition in the disc midplane \hbox{($I^+ = I^-$)}, we find
the two boundary conditions
\begin{equation}
F = 0 \ {\rm at}\ z=0 \label{inbound}
\end{equation}
\begin{equation}
F = {{4 \pi}\over{\sqrt 3}} J \  {\rm at} \ z=z_{out} \label{outbound}
\end{equation}
where $z_{out}$ is the outer edge of the disc, which is predefined to
lie at some very low pressure, say comparable to that of the interstellar
medium.
 
In the two-stream approximation the temperature is not obtained by
integrating the temperature gradient (as in the diffusion
approximation),  but is calculated locally from the two variables 
$J$ and $P$ by solving the thermal equilibrium equation
\begin{equation} 
4 \pi \kappa_P(P,T) \rho (P,T) (B(T) - J) = H_{visc}(P) \label{theq}
\end{equation}
where $H_{visc}$ is the viscous heating rate per unit volume,
\begin{equation}  
H_{visc} = 1.5 \alpha \Omega P \quad . \label{hvisc}
\end{equation}

From our solutions of the vertical disc structure using the two-stream
approximation as described above, we will find that the thermal structure
of the outer layers of the disc
(as defined by the temperature of the solution as a function of
pressure) no longer depends on radiative transfer effects. This can be
understood as follows. At low optical depth and low 
pressure, the flux $F$ is nearly constant, being 
determined by the dissipation in the underlying layers. Furthermore, 
we find that the outer boundary condition (\ref{outbound}) is not only
satisfied at the exact outer boundary, but is a very
good approximation throughout the optically thin region.
In this case the temperature equilibrium equation can be simplified to
\begin{equation}
4 \pi \kappa_P(P,T) \rho (P,T) \left({{\sigma_{SB}
T^4}\over {\pi}} - {{{\sqrt 3} \sigma_{SB} T_{eff}^4}\over{4 \pi}} \right)  =
H_{visc}(P)
\label{theqs}
\end{equation}
where $\sigma_{SB}$ is the Stefan-Boltzman constant.
Using this equation, we can solve for the heating/cooling equilibrium as a
function of the local variables pressure and temperature in
the outer layers of the disc for given values of the central mass,
mass accretion rate (which is equivalent to effective temperature)
and radius without having to consider the vertical disc structure.

{\subsection{Opacities} \label{opsec}}

Previous applications of the grey two-stream method to accretion disc
vertical structure have not always taken the correct type of opacity
average in the moment equations, and have used the Rosseland mean absorption 
coefficient rather than the Planck mean in equation \ref{dFdz} (ASSW).    
The Rosseland mean is basically a measure of how easily
radiation can escape. It is designed to give the correct physical
result when the diffusion approximation is used, and is thought to be
a good approximation of the flux mean, hence its appearance in
equation \ref{dJdz}. It is mainly determined by how much of the
spectral range has a relatively low opacity. Our numerical results
confirm the assumption underlying the simplification made in previous
works, in that the
structure of the main body of the accretion disc is in fact extremely
insensitive to the mean opacity in equation \ref{dFdz}.
 
However, there is a problem with using the Rosseland mean when considering
the structure of the optically thin regions that we are interested in
in this paper. The optically thin cooling rate $C$ is given by definition as
\begin{equation}
C \equiv 4 \pi \int \kappa_\nu B_\nu (T) d \nu 
\equiv 4 \pi \kappa_P B(T)  \label{crate}
\end{equation} 
with $\kappa_P$ the Planck mean opacity. If the cooling is mainly due
to lines or other features with a high absorption
coefficient over a narrow frequency range, the Planck mean can be
much higher than the Rosseland mean, and the use of the Rosseland mean
in equation \ref{theq} severely underestimates the cooling
rate. Comparing Rosseland and Planck mean opacities from the Opacity
Project the difference can exceed a factor of $10^3$, and
even for a continuum process like free-free absorption the
Planck mean is about 40 times higher than the Rosseland mean (Rybicki
\& Lightman 1979). In our application, it is therefore essential to
use the Planck mean in equations \ref{dFdz}, \ref{theq} and
\ref{theqs}.

The Rosseland mean opacities we use are calculated with a
program provided by R. Wehrse (private communication), and the Planck mean
opacities are based on opacities from the Opacity
Project (OP, Seaton et al. 1993) which are the state of the art in
including as many bound-bound and bound-free processes as feasible.
The reason we use different sources for the two mean opacities is as
follows. The opacity program by
Wehrse can calculate the opacities from physical principles
over the entire range of pressures and temperatures needed for our
accretion disc models, whereas the OP tables cover only a part
of it. On the other hand, the program does not consider many
spectral lines that are quite important for optically thin cooling. 
The Rosseland mean opacities from these two sources 
are virtually indistinguishable in the $(P,T)$ region where both are
available, but the Planck mean resulting from Wehrse's program is much
lower than that from the OP data because many lines are not included. 

Using the OP mean opacities for accretion discs two problems are
encountered. The first is that the range of density and temperature
for which these opacities have been computed do not overlap the entire
regime for which they are needed, so that extrapolation to
low density is necessary. 
The second one is that they are calculated for LTE, which
will no longer be valid in the low density optically thin outer
layers of the disc. In the low density limit we have
\begin{equation}
4 \kappa_P \rho \sigma_{SB} T^4 \equiv \Lambda n_H^2 
\label{lambda}
\end{equation} 
where $\Lambda$ is the ``cooling function'' calculated by many authors
(e.g. Raymond et al. 1976, Sutherland \& Dopita 1993). 
We therefore extrapolate the OP
opacities to low densities assuming that below the lowest density
available on the OP grid $\kappa_P$ is reduced proportional to the
density. The ``cooling function'' that can be derived from the OP Planck
mean opacities using equation \ref{lambda} is mostly within a factor of 
a few from the detailed cooling functions quoted above, except for 
low temperatures below $1.5 \times 10^4$K where it is significantly higher.  
In spite of these difficulties with the OP opacities, we still
consider them an improvement over using the severely underestimated
Planck mean opacities as calculated with Wehrse's program.

\section{Results}

\subsection{Vertical structure models}

Here we first describe some results of our vertical structure
calculations with the grey two-stream method to be able to put the results of
our surface layer models into context.
We solve the basic set of differential equations 
\ref{hstat}, \ref{dJdz}, \ref{dFdz} with boundary conditions
\ref{inbound}, \ref{outbound} and $T=T_c$ at $z=0$, 
together with the auxiliary equations
\ref{eqstate}, \ref{theq} and \ref{hvisc}, 
for the four variables $P$, $T$, $J$ and $F$ 
as a function of z.  
The integration is started in the midplane (z=0),
and proceeds outwards. In previous applications of the
two-stream grey method the integration was usually started at the
surface, but because we want to find the 
point where the thermal equilibrium equation does not have a solution
any more, we can not use this method here. 

Rather than solving for the
disc structure by iterating on the disc height for a given mass
accretion rate (or flux at the surface boundary), we solve for a self 
consistent disc structure by iterating on the central density for a
given central temperature $T_c$, and the mass accretion rate and surface
density are determined by the model. 
The density at z=0 is iterated until the outer
boundary condition (\ref{outbound}) is fulfilled at a pressure where the
optical depth is already very low (say $10^{-3}$). This solution
determines the main structure of the disc. We then do one more
integration outwards with the midplane density and temperature from
the converged solution, but now the integration does not stop at the
pressure previously defined as the outer boundary, but continues
outwards until the thermal equilibrium equation does not have a
solution any more. It was carefully verified that the instability
point obtained in this way does not depend on the choice of
pressure at the outer boundary in the initial disc structure
iterations as long as it lies at sufficiently low optical depth and
pressure. Below we present some results for an accretion disc with
$\alpha =0.3$ around a one solar mass compact object.
\begin{figure}
\mbox{\psfig{figure=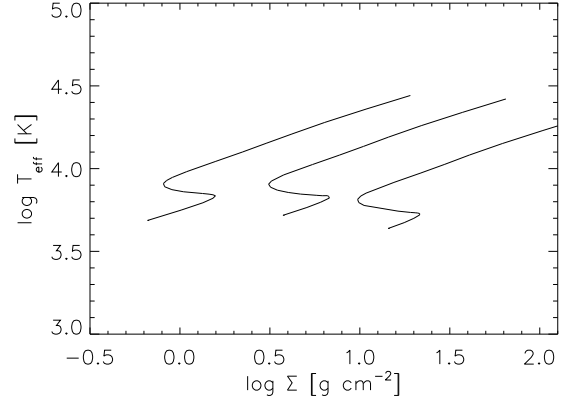,height=6.cm}}
\caption{The relation between 
the surface density and the effective
temperature of an accretion disc around a 1 \Msun \ object that we
obtain from our vertical structure calculations, for radii $10^9$,
$3.5 \times 10^9$ and $10^{10}$ cm.}
\label{sig-teff}
\end{figure}
In Figure \ref{sig-teff} we show the relation 
between surface density and effective
temperature (the ``S-curves'') for radii $10^{10},\ 3.5 \times 10^9$
 and $10^9$ cm. 
The lowest $T_{eff}$ models have a $\tau_R$ of a
few tenths, and are therefore already unreliable since the radiation
field is no longer approximately a Planck spectrum, and the effective
mean opacity will be very different from the Rosseland mean (Mineshige
\& Wood 1990). We
therefore do not extend our models to lower optical depth discs. These
curves reproduce reasonably well previous calculations carried out by
other investigators (e.g. Canizzo \& Wheeler 1984 ) in the diffusion
approximation. 
\begin{figure}
\mbox{\psfig{figure=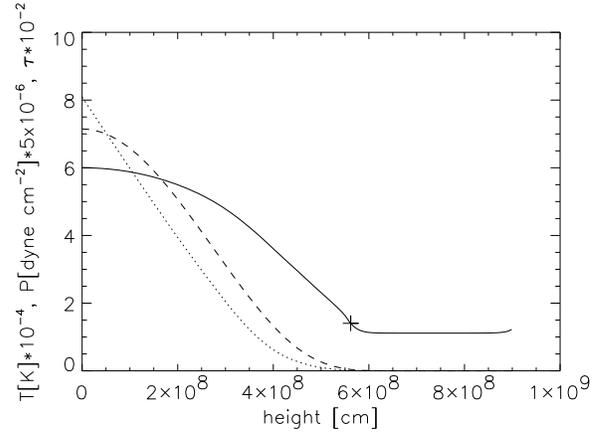,height=6.cm}}
\caption{An example of the vertical structure of an accretion
disc. Shown are the temperature (solid), pressure (dashed) and
Rosseland mean optical depth (dotted)
as a function of height above the disc plane for an accretion disc
around a 1 \Msun white dwarf, at a radius of $10^{10}$ cm, and for a mass
accretion rate of $10^{-9}$ \Msun yr$^{-1}$. }
\label{zstruc}
\end{figure} 
In Figure \ref{zstruc}
 the vertical structure of a disc model at $R = 10^{10}$\ cm
with an accretion rate of 
$\Mdot = 10^{-9} \Msun {\rm yr}^{-1}$ is shown in detail. Plotted are
the temperature, pressure and Rosseland mean optical depth as a
function of  height above the disc mid-plane. The cross on the
temperature  curve indicates
the point where the optical depth is 1. 
The most pronounced feature is the long
temperature plateau at low optical depth, which is explained in the
next section. The temperature only starts to rise significantly,
signalling the approach of the instability point, at a very low
optical depth of less than $10^{-9}$. Note that there is no
hydrostatic corona in this model: the temperature changes
continuously, and at the pressure where
 the low-temperature equilibrium is lost there
is no high-temperature equilibrium available to the gas. This is
similar to the case when only free-free processes are considered
(SWW). 
\begin{figure}
\mbox{\psfig{figure=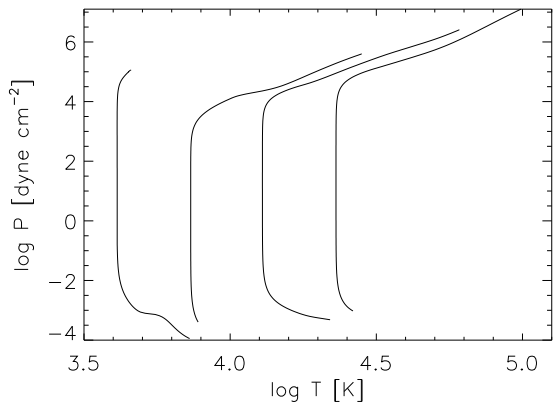,height=6.cm}}
\caption{The pressure as a function 
of temperature from our vertical
structure calculations, for a disc around a 1 \Msun white dwarf. It
represents a series of models at a radius of $3.5 \times 10^9$ cm,
with $\Mdot_{disc}$ equal to $10^{14} , 10^{15} , 10^{16}$ and $10^{17}$
\hbox{g s$^{-1}$} from left to right}
\label{vs_eq}
\end{figure} 
Figure \ref{vs_eq} shows the relation between temperature and pressure for a
number of models with increasing mass transfer rate
 [$\Mdot_{disc} = 10^{14} , 10^{15} , 10^{16}$ and $ 10^{17}$
\hbox{g s$^{-1}$}  ]
at $R=3.5 \times 10^{9}$\ cm.
All structures have the same general morphology:
pressure and temperature drop simultaneously with increasing height,
until the point at which $\tau_R = 1$ is reached. At this point the
temperature becomes constant at a value slightly below the effective
temperature, while the pressure drops by about 6 orders of
magnitude. At very low pressure the temperature starts to rise again
until the thermal instability sets in. 

\subsection{Simple surface layer models}

\begin{figure}
\mbox{\psfig{figure=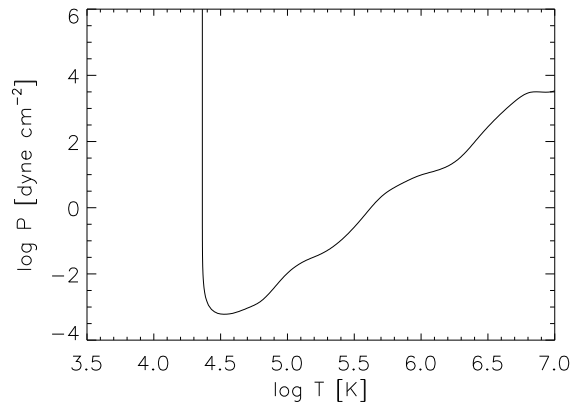,height=6.cm}}
\caption{The thermal equilibrium curve 
in the $(P,T)$ plane that is
calculated from equation \ref{theqs}. The parameters are the same as
for Figure \ref{vs_eq}, except that only the results for $\Mdot_{disc} =
10^{17}$ g s$^{-1}$  is shown.}
\label{simpeq}
\end{figure}

In this section we will show how the simplified thermal equilibrium
equation \ref{theqs} explains the behaviour of the full disc solution
at low optical depth. In Figure \ref{simpeq} we show the solution of equation 
\ref{theqs} in the $T,P$ plane for $R = 3.5 \times 10^{9}$\ cm and  
$\Mdot = 10^{17} {\rm gr s}^{-1}$ .  At high
pressures there are two equilibrium temperatures. The high temperature
solution mostly lies on an equilibrium curve with a positive
${d~log P \over d~log T}$ gradient, and is unstable since after a
 temperature perturbation at constant
pressure (dictated by the hydrostatic equilibrium), the gas will tend
to move farther away from the equilibrium line. On the other hand, the
low temperature
equilibrium is stable. For low temperatures, the temperature is almost
independent of the pressure. This is because $H/\kappa \rho <<
\sigma_{SB} T_{eff}^4$, so that the equilibrium temperature is given
to high precision by $T = ({{\sqrt 3}\over 4})^{0.25} T_{eff}$. Since at low
densities $\kappa \propto \rho$, for low enough pressures the
heating term becomes comparable to the absorption and emission terms,
and the equilibrium temperature rises. The minimum in the thermal
equilibrium curve defines the lowest pressure at which thermal
equilibrium is possible.
\begin{figure} 
\mbox{\psfig{figure=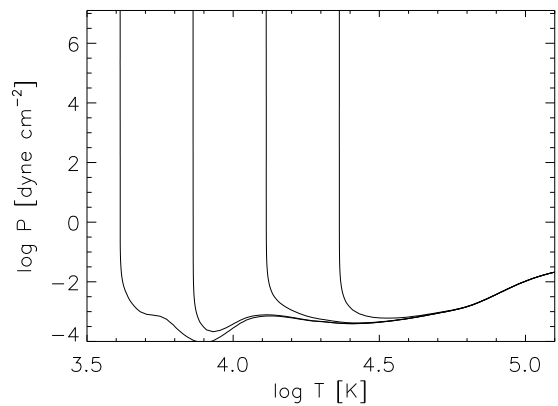,height=6.cm}}
\caption{A close-up of the thermal 
equilibrium curves resulting from
equation \ref{theqs}, on the same scale as Figure \ref{vs_eq}. 
The curves are for 
$\Mdot_{disc} = 10^{14} , 10^{15} , 10^{16}$ and $10^{17}$
 g s$^{-1}$ from left to right respectively.}
\label{simpeq_cu}
\end{figure}  
In Figure \ref{simpeq_cu}
 we show a close-up of the solution of equation \ref{theqs}
on the same scale as in Figure \ref{vs_eq}, for $R = 3.5 \times 10^{9}$\ cm and 
$\Mdot = 10^{14}, 10^{15}, 10^{16}$ and $10^{17}\ {\rm gr s}^{-1}$.  
Comparing Figure \ref{simpeq_cu} with Figure \ref{vs_eq}, it is
obvious that the disc models just follow the thermal equilibrium curve
as soon as the optical depth becomes small. Therefore, to estimate the
temperature and pressure at the instability point we do not have to 
calculate the entire disc model, but only the equilibrium curves.

From Figure \ref{simpeq} it is clear that these vertical structure models
never form a significant static hot corona. We note, however, that for
a narrow range of pressure
and temperature close to the global minimum in the equilibrium curves, there
 is a second higher temperature branch of 
negative ${d~log P \over d~log T}$ where another stable equilibrium solution
appears to be possible. 
Thus, we see from the close-up of Figure \ref{simpeq_cu}, 
that for a very narrow range of $\log \Mdot$ between 15 and 16  
the equilibrium curve has two minima, the
higher temperature one ( $2 \times 10^4$ K)  being slightly lower in pressure
than the lower temperature ($\sim 10^4$ K)
one. The difference in pressure between the two minima is so small,
however, that this cannot be considered a significant effect. Anyway,
our cooling description is not likely to be accurate to such fine
detail because of uncertainties in the opacities.

\section{Detailed surface layer models}

Clearly the description of the radiative heating and cooling processes 
in terms of a single (very uncertain) mean opacity is not very satisfactory.
The adequacy of using mean opacities for optically thin discs and
accretion disc atmospheres is also addressed in detail by
Mineshige \& Wood 1990, who found that models with a frequency
dependent opacity can lead to significantly different results than
those with mean opacities.
Furthermore, the model above neglects the possible effect of a dilute
external radiation field from the centre of the accretion disc or the
surface of the accreting star when one is present. This external
radiation field can be quite important, as has already be shown in
other models (e.g. Begelman \& McKee and Shields 1983, Idan \& Shaviv
1996).

The previous section has shown that, in the outer very optically thin 
layers of the accretion disc where the thermal instability occurs, the
temperature as a function of pressure is given to high accuracy by the
thermal equilibrium curve in the $(P,T)$ plane of an optically thin
gas irradiated by the underlying disc (equation \ref{theqs}). 
This result allows us to construct a more
accurate description of the low optical depth region by calculating
such thermal equilibrium curves for optically thin irradiated gas in
more detail by using a different tool. It turns out that the pressure, 
temperature and optical depth in these outer layers is such that it 
is possible to calculate the thermal equilibrium curves with a  
photoionisation equilibrium code, with an extra heating term added 
to describe the viscous energy dissipation. We have used the code
MAPPINGS (Sutherland \& Dopita 1993). As all other photoionisation codes,
this code is designed to work for a low density environment, and does
not make any assumptions like LTE. The downside is that such codes
become less accurate when the density becomes very high, because 
collisional de-excitation for species other than hydrogen and helium 
is not included, which could quench some of the emission
lines. This limit on the density means that we can
not consider the innermost parts of accretion discs around neutron
stars or stellar mass black holes because the pressures and densities
at the point where the optical depth becomes small are too high for
the photionisation equilibrium code. 

Thus, we consider the thermal equilibrium of an optically thin parcel
of gas heated by viscous
heating (equation \ref{hvisc}) and irradiated by a radiation field
with two components: an undiluted blackbody with a temperature equal
to the effective temperature of the underlying accretion disc, and a
diluted, generally much hotter blackbody coming from the central parts
of the accretion flow. The fact that the disc models above indicate that the
instability point lies at extremely low ($\sim 10^{-7} - 10^{-10}$) Rosseland
mean optical depth justifies our neglect of radiative transfer
effects. We have also neglected the possible effect of shadowing of the
radiation coming from the inner parts of the accretion flow by the
disc at radii smaller than the one considered.  

In terms of input parameters, the surface layer models are a
function of the central mass $M$, the mass flow rate through the disc
$\Mdot_{disc}$, the radius $r$ in the disc considered, 
the viscosity parameter $\alpha$, and the external radiation field.

\subsection{Application to Cataclysmic Variable Discs}

For the external irradiation of a CV disc we take the 
radiation field expected from the boundary layer. We assume that
for accretion rates below $10^{16}$ g s$^{-1}$
the boundary layer is in a hot ($T \sim T_{virial} \sim 10^8$ K)
state (Warner 1995), leading to highly diluted, hard irradiating
spectrum. For $\Mdot_{disc} \ge 10^{16}$g s$^{-1}$, the
boundary layer is optically thick, and we take the radiating area to
be $2 \pi R_{WD} h_{BL}$, with $R_{WD}$ the white dwarf radius and 
$h_{BL}= 0.25 R_{WD}$. In all cases we
assume that the total luminosity of the boundary layer is 
\begin{equation}
L_{BL} = { 1 \over 2}{{G M \Mdot}\over{R_{WD}}} \quad .
\end{equation}
That is, we assume that the heating is what is expected for a steady disc
corresponding to the given $\Mdot$.
\begin{figure}
\mbox{\psfig{figure=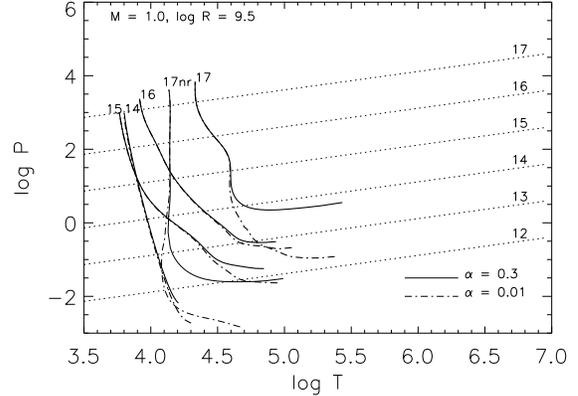,height=6.cm}}
\caption{The thermal equilibrium curves for the viscously heated
outer layers of a CV accretion disc we obtain from the detailed
photoionisation calculations. Equilibrium curves are shown for two values
of the viscosity parameter, $\alpha = 0.3$ and 0.01. Each curve is
marked with the logarithm of the mass flow rate through the disc in
\hbox{g s$^{-1}$}. The curve marked 17nr is for the same parameters as
that marked 17, except that the irradiation from the boundary layer
is not included. The dotted lines are lines of equal total mass
loss from the cool disc by evaporation as calculated by equation
\ref{Mdwind}, and are also labelled with the logarithm of the mass
flow rate. }
\label{cv95}
\end{figure}   
Figure \ref{cv95} shows the result of the thermal equilibrium analysis of  
the outer layers of a CV disc at $r= 3.5 \times 10^9$ cm, for a range of 
mass flow rates through the disc ($\Mdot _{disc}$) ranging from $10^{14}$ to 
$10^{17}$ g s$^{-1}$, and two choices for the viscosity parameter 
$\alpha$, 0.3 and 0.01. We show only the part of the equilibrium curve
before or slightly beyond the minimum, since for higher temperatures
only unstable equilibria exist.

The irradiation flux, and its spectrum, determines 
the pressure and the temperature
at which thermal instability sets in, but does not change the
overall qualitative behaviour of the thermal equilibrium curves.
To demonstrate the former, we also show two models with the external
radiation field turned off. These are the curves marked 17nr in Figure
6. We can see that, to good approximation,
the effect of this on the equilibrium curve is the same as reducing 
the mass accretion rate in the models including external radiation by 
a factor of $10^2$.

The general morphology of the curves is quite similar to Figure 
\ref{simpeq_cu}. Most directly comparable to Figure \ref{simpeq} 
is the non-irradiated curve
17nr (solid), which has exactly the same parameters
($\alpha=0.3$, \hbox{$\Mdot = 10^{17}$ g s$^{-1}$}). 
In detail there is some difference. The temperature over most of the
atmosphere is lower than that derived using a mean opacity,
showing that the assumptions (mainly LTE) underlying the mean opacity
calculations are not valid in the low density outer layers. The new
minimum lies at a similar temperature ($\log T \sim 4.5$), but at a
10 times higher pressure. The inclusion of the external radiation
field raises the critical pressure by an additional factor of 100. 
The detailed models are also similar to
those in the previous section in that we never find hydrostatic
coronae, since there is no hot equilibrium phase available to the gas
once the pressure has dropped below the minimum.

Following Czerny \& King 1989, we estimate the mass loss rate 
that can result from the thermal instability from the pressure and
temperature at the instability point (see also SWW):
\begin{equation}
\Mdot_w(r) \sim 2 \pi r^2 \rho_{inst} c_{s,inst} \label{Mdwind}
\end{equation}
where $\rho_{inst}$ and $c_{s,inst}$ are the density and sound speed
at the minimum in the thermal equilibrium curve. We consider this
estimate to be an upper
limit because the real mass loss rate is determined by $\rho c_s$ at
the critical point where $v = c_s$. This critical point must lie at a
higher z than the thermal instability point where the outflow starts
to be driven, and we know that $\rho c_s \propto P/{\sqrt T}$ is a
decreasing function of height.    

The dotted lines in Figure \ref{cv95} are lines of
equal mass loss as defined by equation \ref{Mdwind}, labelled with the
logarithm of the mass loss rate from the cool disc. We see that the
minimum in the thermal equilibrium curves occur at a temperature and
pressure that imply an evaporation rate that is
always less than $10^{-2}$ times the accretion
rate, with the largest relative mass loss occurring for the lowest
accretion rates. We also find that the relative importance of
evaporation (as defined by the ratio $\Mdot _w / \Mdot _{disc}$) is
not very dependent on radius being about 3 times higher at $10^{10}$ than
at $10^{9}$ cm.

On the basis of the above calculations, we can conclude that mass loss
by evaporation due to the mechanism considered here will occur, 
but cannot be very important for the overall structure of ${\it steady}$
CV discs.
One may ask if the mechanism could be more effective in non steady discs,
such as those found in dwarf novae, and perhaps be responsible for the central
``holes'' that appear to develop in some systems after an outburst. According
to the limit cycle model for dwarf novae, the mass transfer rate through 
the disc oscillates between the values appropriate to the upper and
the lower branch of the equilibrium S curves shown in Figure
\ref{sig-teff} (Osaki 1996), in general 
agreement with observations (Warner 1995). 
After a dwarf nova outburst the mass transfer rate through the disc
could  drop by
a factor of $\sim 10^3$ from say $\Mdot _{disc}= 10^{17}$ to
$\Mdot _{disc}=10^{14}$ g s$^{-1}$,
but the disc may continue to be heated by a luminosity that is
significantly  higher than the value expected for a steady disc at
the  lower accretion rate.  The irradiation luminosity could be by
soft photons from the white dwarf which  has been accretion heated by
the  immediately preceding outburst, and accumulated effects of 
previous outbursts. Enhanced heating of the outer disc may also
occur  at the end of an outburst when the  luminosity of the inner
regions  of the disc far exceeds
the luminosity expected for steady disc at the mass transfer rate rate
appropriate to the lower branch of the S curve.
  
\begin{figure}
 \mbox{\psfig{figure=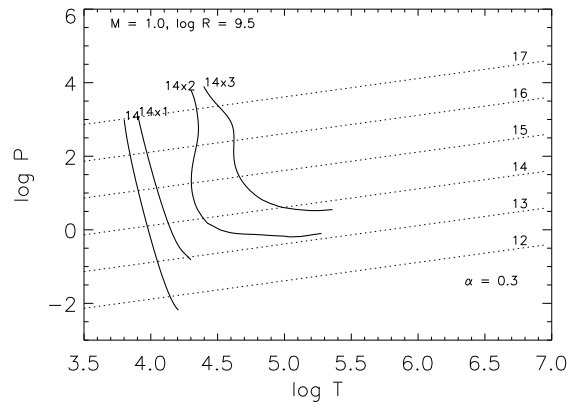,height=6.cm}}
\caption{The thermal equilibrium curves for a disc with the same
parameters as in Figure \ref{cv95}, for a mass flow rate of $10^{14}$ \hbox{g
s$^{-1}$} through the disc, but with increased irradiation from the boundary
layer, as could happen during an outburst phase. To model the
increased irradiation, the mass flow rate in the boundary
layer has been increased by a factors of 10, 100 and 1000 relative to
that in the disc at $3.5 \times 10^9$ cm.}
\label{cv95extra}
\end{figure}

In order to investigate the effects of non-steady heating, we
carried  out an additional series of calculations for the model with
$\Mdot _{disc}=10^{14}$  g s$^{-1}$ where the heating  was enhanced to
correspond  to mass transfer rates near the inner edge of the disc of 
$\Mdot _{disc}=10^{15},10^{16}$ and $10^{17}$  g s$^{-1}$
respectively (Figure \ref{cv95extra}).  At the upper end of this
range, the heating
is  mainly by  UV photons at the appropriate
black body temperature, and disc evaporation becomes very
effective.    These calculations show that non steady heating, for
instance  due to radiation from the heated white dwarf, can increase
disc evaporation significantly. However, an increase in irradiation by
a factor of at least $10^3$ over the steady case is necessary for the
evaporation rate to become comparable to the mass flow rate through
the disc. 

\subsection{Application to stellar mass Black Hole discs}

\begin{figure}
\mbox{\psfig{figure=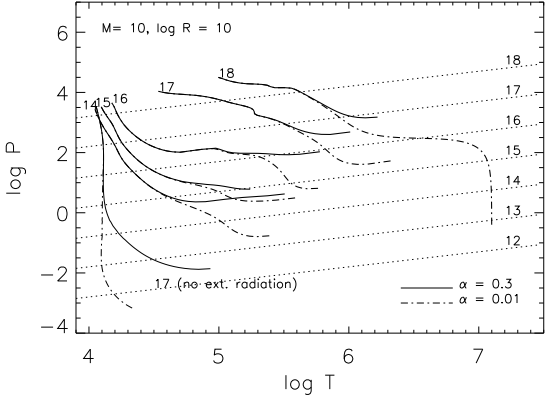,height=6.cm}}
\caption{The same as Figure \ref{cv95}, but now for an accretion disc around a
10 \Msun black hole at a radius of $10^{10}$ cm. Note the much
stronger effect of the irradiation.}
\label{bh10}
\end{figure}

The main difference between the thermal equilibrium in the outer
layers of an accretion disc around a stellar mass black hole and a CV
is the role played by the external hard radiation field
(e.g. Begelman, Shields \& McKee 1983, Idan \& Shaviv 1996). As shown
above, for CV's this external radiation field has a mild effect on the 
steady state structure, but
for black hole discs the radiation field completely changes the
behaviour of the equilibrium curves. For discs around a stellar mass
black hole, we assume that the temperature of the external
radiation field is
\begin{equation}
T_{bb} = T_{d,eff} (r_{max})  \label{tbbc}
\end{equation}
where $T_{d,eff}$ is the effective temperature of the disc and  
$r_{max}$ is the radius of the disc where most energy is dissipated,
at 10 Schwarzschild radii. The dilution factor $\epsilon$ is
\begin{equation}
\epsilon = \eta {{\pi r_{in}^2}\over{4 \pi r^2}} \label{dilf}
\end{equation}
with $\eta$ a geometrical correction factor for the fact that the
central continuum source is likely to be beamed perpendicular to the
disc. We take $\eta$ to be 0.1. In Figure \ref{bh10} we show the
thermal equilibrium curves for the outer layers of a disc at a radius
of $10^{10}$ cm around a 10 \Msun black hole, parameters 
 appropriate for BHSXT's.

First we demonstrate the importance of the external radiation field by
comparing a model with $\Mdot_{disc} = 10^{17}$ g s$^{-1}$ with (the
thick lines labelled with 17) and without (thin lines) radiation. 
The presence of the radiation field increases the temperature at the
instability point from $6 \times 10^4$ to $7 \times 10^5$ K and the
pressure from $10^{-2}$ to $3 \times 10^2$ dyne cm$^{-2}$ for the
$\alpha = 0.3$ curve. According to our mass loss estimate equation
\ref{Mdwind}, this implies a factor of $10^4$ increase in mass loss
from the cool disc relative to the non-irradiated case. The models
shown in Figure \ref{bh10} also do not exhibit a real corona in the
sense that a sudden transition to high temperatures occurs, although
for the higher mass loss rates the temperature goes up to a million
degrees before becoming unstable. We note that for larger radii and
higher mass accretion rates a temperature discontinuity does form, with
the temperature jumping from a few times $10^4$K to about $10^6$K, the
latter corresponding to a well known equilibrium state of an
X-ray irradiated low density gas (Krolik, McKee \& Tarter 1981). 
The gas can remain in
the hot state while the pressure drops by another factor of ten before
becoming unstable. 

It can be seen from figure \ref{bh10} that
for the highest accretion rate considered ($10^{18}$ g s$^{-1}$) and
$\alpha = 0.01$, the disc atmosphere does not become unstable at low
presssures, but rather assumes a constant temperature as the pressure
drops. This is because in this case the radiation field is strong
enough to compensate for the viscous heating by inverse Compton
cooling. Both the viscous heating rate and the Compton cooling rate
are proportional to the pressure, so that the equilibrium remains valid as
the pressure drops, and the atmosphere is stable everywhere. However,
since the viscous heating rate is proportional to the disc angular
velocity $\Omega \propto r^{-1.5}$, and the Compton cooling
proportional to the energy density of the radiation field $U_{rad}
\propto r^{-2}$, the cooling decreases faster with radius than the
heating, and beyond a certain radius the thermal instability
returns. In our case, we find that models with an accretion rate of 
 $10^{18}$ g s$^{-1}$ and $\alpha = 0.01$ become unstable for
$\log R \gtorder 10.5$. These results are consistent with 
those of Czerny \& King 1989, who studied the stabilization of an
accretion disc atmosphere by Compton cooling
while neglecting other radiative cooling and heating processes. 
Czerny \& King considered only radiation directly from the underlying disc. 
Because we take the external irradiation into account, 
we find more effective stabilization at larger radii than Czerny \& King did.

We point out that the models for Compton heated winds
from accretion discs of Begelman, McKee and Shields (1983) did not
include viscous heating. Most of the
effect of these Compton heated winds come from the outer radii of
the disc where the virial temperature becomes comparable to the
Compton temperature. In our calculations the region where Compton
cooling can compensate for viscous heating does not extend out far
enough to reach $T_{comp} \sim T_{vir}$, so that the regime considered 
in the Compton heated wind models is never reached. 
Thus, the Compton heated wind models require
that $\alpha << 0.01$ in the corona in order to be valid.

\begin{figure}
\mbox{\psfig{figure=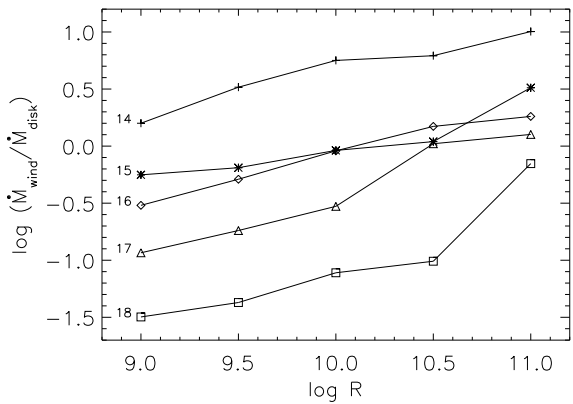,height=6.cm}}
\caption{This plot illustrated the relative importance of viscous
evaporation, by showing the ratio of the evaporative mass loss rate
(equation \ref{Mdwind}) and the mass flow rate through the disc, for
a disc around 10 \Msun black hole with $\alpha = 0.3$. For low
$\Mdot_{disc}$ the disc can
be completely evaporated in the range of radii considered. As
$\Mdot_{disc}$  increases, evaporation only becomes important at larger
disc radii.}
\label{mdrel}
\end{figure}

Figure \ref{bh10} also shows that if $\alpha \sim 0.3$, the vertical mass loss
from the cool disc could become comparable or greater than the mass flow
through the disc for $\Mdot_{disc} \ltorder 10^{16}$ g s$^{-1}$,
implying that steady, cold discs may not exist around stellar mass
black holes in the low $\Mdot$ regime if the dissipation rate in the
 outer layers is high
($\alpha \gtorder 0.1$). We have summarized our results for the
relative mass flow out of the cool disc ($\Mdot_{w}/ \Mdot_{disc}$)
for different radii in Figure \ref{mdrel}. Our results indicate that
disc evaporation by a combination of viscous heating and irradiation
is more important in the {\em outer} parts of a steady disc than in
the inner parts. Only for the lowest accretion rate of $10^{14}$ g
s$^{-1}$ is the entire range of radii we consider susceptible to
complete evaporation, implying that a steady state cool disc may not
exist beyond a radius of $10^9$ cm. However, the evaporation could remain a
significant effect at at least the 10 \% level for all accretion rates
considered. We re-emphasize that this conclusion does depend upon the
effectiveness of viscous heating ($\alpha$) in the atmosphere, and that for
smaller $\alpha$ the evaporation becomes important only for lower mass
accretion rates and larger disc radii, e.g. for $\alpha = 0.01$ the
accretion rate and evaporation rate become comparable for R =
$10^{11}$ cm and $\Mdot_{disc} = 10^{14}$ g s$^{-1}$.     

The somewhat complex behaviour (crossing over) of the
curves at log R = 10.5 in Figure \ref{mdrel} is due to a switch of 
the equilibrium state at which the lowest stable pressure occurs,
between  the cool
equilibrium at a few times $10^4$ K and the hot equilibrium
for X-ray irradiated gas at about $10^6$ K.

\section{Discussion}

We have shown that mass loss can occur from
the surface of an accretion disc by evaporation due to thermal
instability for a variety of conditions. 

For accretion discs in cataclysmic variables, radiative heating by 
photons from smaller radii in the form of
hard X-rays (from an optically thin boundary layer) or $\sim 10^5$K
black body photons (from an optically thick boundary layer or the heated white 
dwarf surface), plays an important role in determining the
mass loss rate due to atmospheric thermal instability. Under
steady state conditions, the mass loss rate is expected to be
less than a maximum of a few percent of the mass transfer rate through 
the disc, with the maximum being reached in low $\Mdot$ discs. However,
in non steady discs, mass loss from thermal instability could be a 
larger fraction of the mass transfer rate through the disc.  Another
effect that could significantly increase the evaporation rate is extra
heating in the atmosphere, as would be expected if a significant
buoyant flux of magnetic field is escaping from the disc and is
dissipated in the atmosphere, as has been suggested by some authors
(Tout \& Pringle 1992, but see also Stone et al. 1996).  
The thermal instability may therefore provide an explanation for the
UV delay seen in some dwarf novae, which has been interpreted
as evidence for the existence of "holes" in the disc during the low $\Mdot$
phase of the dwarf nova cycle, but other effects that come into play
once the instability has led to a significant corona are probably a
better candidate.
  
An example of this is the work of Meyer \& Meyer Hofmeister 1994 (MMH) and 
Liu et al. 1996, who found that evaporation from a CV disc 
can easily lead to significant mass loss. These models do not consider
the stability of a cold disc by itself, but rather consider the situation in
which the disc is sandwiched between a corona, and calculate
the structure of such a disc-corona combination self-consistently. The
evaporation in their case is due to thermal conduction from the
viscously heated corona into the cold disc. They also take the
dynamics of the evaporative outflow into account, an ingredient that
is still missing from our exploratory calculations. 
Our results indicate that a cold disc by itself is indeed unstable,
and that the formation of a corona is inevitable. Thus,
evaporation by the mechanism modelled here may play a role by
providing seed material for a corona, which may then develop into a 
coronal siphon flow in the manner outlined by MMH.
We do treat the thermal equilibrium of the accretion disc outer layers 
in more detail than MMH, and show that the effect of the radiation
field, which was not included by MMH, can be quite significant.
We find that in our case the heating in the disc atmosphere 
is dominated by absorption of radiation over many pressure 
scale heights (even in the case
of no radiation from the centre of the accretion flow), and not 
viscous heating as modelled by MMH. We expect that for discs around
black holes or neutron stars the irradiation effect is so strong that
the disc evaporation picture developed by MMH for CV discs will have to
be modified to include the irradiation. 

Disc atmospheres around stellar mass black holes develop a thermal 
instability at much higher pressure than discs around white dwarfs due
to the strong irradiation, and significant evaporation by thermal instability
appears to be a real possibility. The importance
of X-ray heating in determining the S curves of BHSXT discs and disc
instability has already been demonstrated by El Khoury and
Wickramasinghe  (1998). Our calculations have
shown that X-ray heating from the inner regions of a black hole disc
can be sufficient 
to drive evaporation at a rate comparable to or larger than the
local mass transfer 
rate through the disc for $\Mdot_{disc} \ltorder 10^{16}$ g s$^{-1}$. 
The evaporation is strongest in the outer regions of 
the disc,and will be quenched by Compton cooling only at very high 
$\Mdot_{disc}$ ($\sim 10^{18}$ g s$^{-1}$) and for 
low $\alpha$ in the accretion disc 
atmosphere. Thus, the thermal instability considered here may play a
role in the transition of thin cool discs to hot advection dominated
discs that has been postulated to explain the absence of a strong
X-ray flux from some accreting black hole candidates (e.g Narayan,
McClintock \& Yi 1996).
   
Our results  could have important implications
for the interpretation of the observations of accretion discs
around  Black Hole Soft X-ray Transients. It is well documented that 
line profiles of H and He lines seen in  BHSXT's are rarely disc -
like.   Rather, phase dependent observations show strongly asymmetric
and split lines which are inconsistent with an origin in a standard
disc (Soria et al. 1998). We draw particular attention to the well
studied system GRO1655-40. The spectral energy distribution during its
1996 outburst appears to evolve in time more like a black body than a
disc (Hynes et al. 1998). The soft X-ray delay seen in this system
(Hameury et al. 1997) also argues for a non standard disc. Based on our
calculations, it would appear that the modelling of the time dependent 
evolution of such systems would require consideration of evaporation,
and accretion from the hot, thick disc that could result from this.
    
\section{Acknowledgements}

We would like to thank Ralph Sutherland for making the photoionisation
code MAPPINGS available to us. Our referee, Ivan Hubeny, has
contributed significantly to making this paper more clear and consistent.

\end{document}